\documentclass[pr, twocolumn, preprintnumbers, showpacs, footnoteadded, superscriptaddress,longbibliography]{revtex4-2}
\pdfoutput=1

\usepackage[T1]{fontenc}
\usepackage{amssymb}
\usepackage{graphicx}
\usepackage{amsmath, amsthm}
\usepackage{slashed}
\usepackage{tensor}
\usepackage{hyperref}
\usepackage{epstopdf}
\usepackage{extarrows}
\usepackage{xcolor}
\usepackage{mathrsfs}
\usepackage{lipsum}
\usepackage{subfigure}
\usepackage{float}
\usepackage{mathtools}

\newtheorem{theorem}{Theorem}

\usepackage{changes}
\usepackage{braket}
\definechangesauthor[name={Per cusse}, color=orange]{per}
\begin{document}
\thispagestyle{empty}
\preprint{\hfill {\small {ICTS-USTC/PCFT-25-57}}}
%<<<<<<<<<<<<< TITLE >>>>>>>>>>>>>>>%
%\title{Pseudospectrum at exceptional points on the black hole}
\title{Exceptional line and pseudospectrum in black hole spectroscopy}	
%\title{The exceptional line in black hole spectroscopy: its topological properties and the pseudospectrum at it}	
%<<<<<<<<<<<<< AUTHOR >>>>>>>>>>>>>>>%
%\author{$^b$}
%
% \email{}

\author{Li-Ming Cao}
\email{caolm@ustc.edu.cn}
\affiliation{Interdisciplinary Center for Theoretical Study and Department of Modern Physics, University of Science and Technology of China, Hefei, Anhui 230026, China}
\affiliation{Peng Huanwu Center for Fundamental Theory, Hefei, Anhui 230026, China}

\author{Ming-Fei Ji}
\email{jimingfei@mail.ustc.edu.cn  (corresponding author)}
\affiliation{Interdisciplinary Center for Theoretical Study and Department of Modern Physics, University of Science and Technology of China, Hefei, Anhui 230026, China}

\author{Liang-Bi Wu}
\email{liangbi@mail.ustc.edu.cn}
% \affiliation{School of Science, Huzhou University, Huzhou, Zhejiang 313000, China}
\affiliation{School of Fundamental Physics and Mathematical Sciences, Hangzhou Institute for Advanced Study, UCAS, Hangzhou 310024, China}
% \affiliation{University of Chinese Academy of Sciences, Beijing 100049, China}

\author{Yu-Sen Zhou}
\email{zhou\_ys@mail.ustc.edu.cn}
\affiliation{Interdisciplinary Center for Theoretical Study and Department of Modern Physics, University of Science and Technology of China, Hefei, Anhui 230026, China}
	
%<<<<<<<<<<<<< ADDRESS >>>>>>>>>>>>>>>%
% \affiliation{${}^a$Interdisciplinary Center for Theoretical Study and Department of Modern Physics, University of Science and Technology of China, Hefei, Anhui 230026, China}

% \affiliation{${}^b$Peng Huanwu Center for Fundamental Theory, Hefei, Anhui 230026, China}

% \affiliation{${}^c$School of Fundamental Physics and Mathematical Sciences, Hangzhou Institute for Advanced Study, UCAS, Hangzhou 310024, China}

% \affiliation{${}^d$University of Chinese Academy of Sciences, Beijing 100049, China}

%<<<<<<<<<<<<< DATE >>>>>>>>>>>>>>>%
\date{\today}

%======================================%
%<<<<<<<<<<<<< ABSTRACT >>>>>>>>>>>>>>>%
%======================================%
\begin{abstract}
We investigate the exceptional points (EPs) and their pseudospectra in black hole perturbation theory. By considering a Gaussian bump modification to the Regge-Wheeler potential with variable amplitude, position, and width parameters, $(\varepsilon,d,\sigma_0)$, a continuous line of EPs (exceptional line, EL) in this three-dimensional parameter space is revealed. Notably, the EL exhibits an anisotropic spectral response: parameters migrating along the EL direction leaves the coalesced QNM spectra nearly unchanged, while moving parameters away from the EL induces the characteristic $\epsilon^{1/2}$ scaling, highlighting the directional nature of spectral instability in exceptional structures. We find that the vorticity $\nu=\pm1/2$ and the Berry phase $\gamma=\pi$ for loops encircling the EL, while $\nu=0$ and $\gamma=0$ for those do not encircle the EL. In the neighborhood of an eigenvalue, through matrix perturbation theory, we prove that the $\epsilon$-pseudospectrum contour size scales as $\epsilon^{1/q}$ at an EP , where $q$ is the order of the largest Jordan block of the Hamiltonian-like operator associated with that eigenvalue, contrasting with the linear $\epsilon$ scaling at non-EPs. Numerical implements confirm this observation, demonstrating enhanced spectral instability at EPs for non-Hermitian systems including black holes. 
\end{abstract}

\maketitle
	
%======================================%
%<<<<<<<<<<< Introduction >>>>>>>>>>>>>%
%======================================%

\section{Introduction}
Upon characterizing the intrinsic oscillatory behavior of perturbed black holes, quasinormal modes (QNMs) manifest themselves as a discrete set of complex frequencies. The oscillation frequency is encoded in the real part, while the decay rate is encoded in the imaginary part. The QNMs not only act as spectral fingerprints that identify black holes, but also offer a powerful probe to test the Kerr hypothesis and explore gravity in the strong-field regime. This program is known as black hole spectroscopy~\cite{Kokkotas:1999bd,Berti:2009kk,Konoplya:2011qq,Berti:2025hly}.

Perturbed black holes are inherently non-Hermitian  systems, a feature common across many branches of physics. Such non-Hermitian systems offer fertile ground for exploring unique topological phenomena absent in Hermitian counterparts~\cite{Moiseyev_2011,El-Ganainy:2018ksn,Ashida:2020dkc,Bergholtz:2019deh}. A defining hallmark of these systems is the appearance of the exceptional points (EPs)~\cite{Heiss_1990,Ryu_2009,Dietz:2010bvm,Xu:2016gkh,doi:10.1126/science.aaf8533,Ding_2016}, where both eigenvalues and their corresponding eigenvectors coalesce. Notably, the adiabatic encircling of EPs generates distinct topological phenomena, including braiding of eigenvalue trajectories and Berry phase accumulation~\cite{Dembowski:2001zz,Dembowski_2004,PhysRevA.85.064103,zhong2018winding},similar phenomenon can also be found in holographic models~\cite{Grozdanov:2019uhi,Ghodrati:2025fah}. EPs naturally emerge in black hole perturbation theory. As demonstrated in~\cite{Motohashi:2024fwt}, the widely studied phenomenon of mode repulsion or avoided crossing in black hole spectroscopy~\cite{Berti:2025hly,Dias:2021yju,Oshita:2025ibu,Lo:2025njp,Takahashi:2025uwo} naturally emerges near EPs, and these EPs can be accessed by introducing a sufficient number of free parameters. This is exemplified by the degenerate QNMs found in massive scalar perturbations of Kerr black holes~\cite{Cavalcante:2024swt,Cavalcante:2024kmy,Cavalcante:2025abr}. Similarly, EPs can be induced by adding a Gaussian bump to the original Regge-Wheeler (RW) effective potential~\cite{Yang:2025dbn}.

A remarkable property of EPs is their singular enhancement of sensitivity to parameter variations, an effect that surpasses anything possible in Hermitian systems~\cite{Ashida:2020dkc}. This enhanced sensitivity has enabled diverse applications, including advanced sensing platforms based on graphene metasurfaces and microtoroid cavities, integrated microresonators operating at three-fold EPs, coupled photonic cavities, and highly sensitive microcavity sensors for nanoparticle detection. Further applications span the detection of mechanical motion, electronics, and two-qubit systems, as detailed in~\cite{Ashida:2020dkc}.

To quantitatively study this enhanced sensitivity in black hole systems, pseudospectrum analysis provides an appropriate framework. Originally developed for non-selfadjoint operators, pseudospectrum analysis measures how eigenvalues respond to perturbations of norm $\epsilon$~\cite{trefethen2020spectra}. Within the hyperboloidal framework, it has been employed to investigate QNM spectrum instability~\cite{trefethen2020spectra,Boyanov:2024fgc, Jaramillo:2020tuu,Destounis:2023ruj,Jaramillo:2021tmt} for the gravity systems. The $\epsilon$-pseudospectrum visually reveals stability through its contour structure in the complex plane, where open contours signify spectrum instability. This method has found broad application across diverse spacetimes~\cite{Jaramillo:2020tuu, Destounis:2021lum,Cao:2024oud,Sarkar:2023rhp,Destounis:2023nmb,Luo:2024dxl,Warnick:2024usx,Arean:2024afl, Cownden:2023dam, Boyanov:2023qqf,Garcia-Farina:2024pdd,Arean:2023ejh,Boyanov:2022ark,Cao:2024sot,Carballo:2025ajx,Chen:2024mon,Siqueira:2025lww,Besson:2024adi,dePaula:2025fqt,Cai:2025irl,Carballo:2024kbk,Cao:2025qws}. For non-Hermitian lattices~\cite{Komis:2022yjk,Ghatak:2024suy}, pseudospectrum has already been used to study the enhanced sensitivity at an EP, where for small $\epsilon$, the boundary of $\epsilon$-pseudospectrum scales as $\epsilon^{1/2}$ at a second-order EP and $\epsilon^{1/3}$ at a third-order EP. It is natural to search for similar behavior of pseudospectrum in black hole spectroscopy or in a more general eigenvalue problem.

In this paper, we modify the Regge-Wheeler potential by a Gaussian bump with amplitude $\varepsilon$, position $d$, and width $\sigma_0$. Although this perturbation is short-range distinguished from the astrophysical long-range perturbation~\cite{Jaramillo:2020tuu}, it provides a reliable model to study the instability of the QNM spectrum due to the environmental effect~\cite{Cheung:2021bol,Yang:2024vor}. We first investigate EPs in the $3$-dimensional parameter space $(\varepsilon,d,\sigma_0)$ of the Gaussian bump-modified Regge-Wheeler potential. By varying the width parameter $\sigma_0$, we find a continuous line of EPs, extending the single EP found in previous studies with fixed width. This line is just an ``exceptional line'' (EL), which widely arises in non-Hermitian systems~\cite{Armitage_2018,Shen_2018,Yang_2019,Wu:2024wlf}. We then examine the topological properties of QNMs associated with this EL by analyzing the winding of eigenvalues and the Berry phase along closed loops in parameter space. Finally, we study the pseudospectrum behavior at these EPs, both theoretically and numerically. Inspired by earlier works for second- and third-order EPs~\cite{Komis:2022yjk,Ghatak:2024suy}, we extend  to a more general case, namely $q$-th-order EPs (see Theorem \ref{theorem1}). For a general finite-dimensional matrix, through matrix perturbation theory, we establish that the $\epsilon$-pseudospectrum contour size scales as $\epsilon^{1/q}$ at an EP, where $q$ is the order of the largest Jordan block, in contrast to the linear $\epsilon$ scaling away from EPs. This conclusion applies to any non-Hermitian system that can be reduced to a finite-dimensional matrix eigenvalue problem, even if the underlying system is itself infinite-dimensional, as in black hole spectroscopy. This theoretical prediction is confirmed by numerical studies on the bump-modified Regge-Wheeler equation.

An introduction to the hyperboloidal framework, a proof of the Theorem \ref{theorem1}, several toy matrix models illustrating pseudospectrum behaviors at EPs, an analysis of sample points of pseudospectra on demonstrating the scalings, and a verification of the anisotropy of spectral instability of EL can be found in the Supplemental Material.

\section{Topological structures of QNMs associated with exceptional line}\label{lineofEP}
In~\cite{Yang:2025dbn}, the authors investigate the QNM resonance arising from adding a Gaussian bump perturbation to the standard Regge-Wheeler potential governing axial (i.e., odd parity) perturbations of the Schwarzschild black hole. Therein, the bump depends on two parameters: the amplitude $\varepsilon$ and the  position $d$. They have found an EP at $\varepsilon=10^{-2.294}\simeq0.005$ and $d=15.698$. Now, we further consider a Gaussian bump with changeable width. To be exact, we consider a master perturbation equation in the time-domain
\begin{eqnarray}\label{tdeq}
    \Big[-\frac{\partial^2}{\partial t^2}+\frac{\partial^2}{\partial r_{\star}^2}-V(r)\Big]\Psi(t,r)=0\, ,
\end{eqnarray}
where $r_{\star}=r+2M\ln(r-2M)$ is the tortoise coordinate and $V(r)$ is the bump-modified Regge-Wheeler potential~\cite{Cheung:2021bol,Yang:2024vor}, namely $V(r)=V_{\text{RW}}(r)+V_{\text{bump}}(r)$ with
\begin{eqnarray}\label{effective_potential}
    V_{\text{RW}}(r)&=&\Big(1-\frac{2M}{r}\Big)\Big(\frac{\ell(\ell+1)}{r^2}-\frac{6M}{r^3}\Big)\, ,\\
    V_{\text{bump}}(r_{\star})&=&\varepsilon \exp\Big[-\frac{(r_{\star}-d)^2}{2\sigma_0^2}\Big]\, ,
\end{eqnarray}
where $\varepsilon$, $d$ and $\sigma_0$ refer to the amplitude, position and width of the bump, respectively. For convenience, we use $2\sigma_0^2$ as the parameter that characterizes the width in the followings.

\begin{figure}[htbp]
    \centering
    \includegraphics[width=0.95\linewidth]{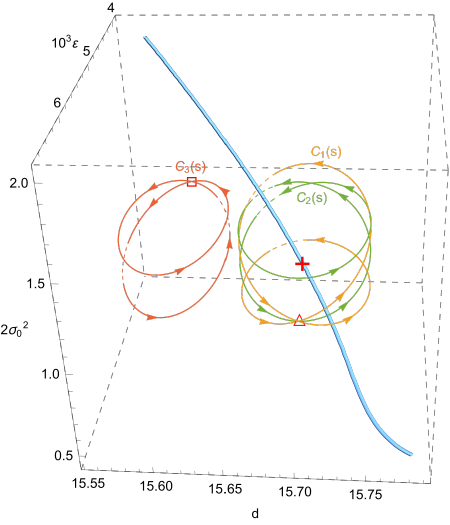}
    \caption{The blue line is the EL in the $3$D parameter space $(\varepsilon,d,2\sigma_0^2)$. The symbol $+$ marks the EP when $2\sigma_0^2=1$. The three closed curves $C_1(s)$ (yellow), $C_2(s)$ (green) and $C_3(s)$ (red). The $\bigtriangleup$ is the point where $C_1(0)$ and $C_2(0)$ coincide, while the $\Box$ represents the point $C_3(0)$.}
    \label{fig:paraloop}
\end{figure}

Within the hyperboloidal framework~\cite{PanossoMacedo:2023qzp,Jaramillo:2020tuu, Gasperin:2021kfv, Besson:2024adi} detailed in the Supplemental Material, the time-domain perturbation equation \eqref{tdeq} can be recast into $\partial_{\tau}U=LU$, where $L/\mathrm{i}$ acts as a Hamiltonian-like operator whose spectra directly yields the QNM spectra. This can be approximated by the eigenvalue problem of a finite-dimensional matrix by discretizing $\sigma$ into Chebyshev-Lobatto grids. We identify the EP where the fundamental mode and the first overtone coalesce, in good agreement with the special case $2\sigma_0^2=1$ considered in~\cite{Yang:2025dbn}. By varying $2\sigma_0^2$, we systematically identify EPs in each constant-$2\sigma_0^2$ plane. These EPs collectively form a continuous line (exceptional line, EL) in the $3$D parameter space $(\varepsilon,d,2\sigma_0^2)$, as shown in Fig. \ref{fig:paraloop}. This demonstrates that EPs exist across a broader range of parameter configurations than previously recognized.

Analysis of Fig. \ref{fig:paraloop} reveals systematic trends: as the Gaussian bump width $\sigma_0$ increases, the amplitude $\varepsilon$ decreases and the bump position moves closer to the event horizon. Along the EL, $2\sigma_0^2$ varies by $300\%$ from $0.5$ to $2$, while $\varepsilon$ changes by approximately $39\%$ from $0.0066$ to $0.0040$, and $d$ changes by only $1.3\%$ from $15.78$ to $15.57$. The corresponding QNM spectra evolve from $0.3671+0.1161\mathrm{i}$ to $0.3648+0.1172\mathrm{i}$, with real and imaginary parts varying by approximately $0.6\%$ and $ 0.9\%$, respectively. This indicates that while $\varepsilon$ and $2\sigma_0^2$ exhibit substantial variation along the EL, $d$ and the corresponding QNM spectra remain relatively stable. This stability of QNM spectra along the EL suggests an anisotropic sensitivity of the spectrum, which we now examine more generally by comparing parameter variations along and away from the EL. As numerically verified in the Supplemental Material, when we vary the parameters in a curve along or parallel to the EL, the two modes change slowly and the leading order of $\omega(s)-\omega(s_0)$ is $\mathcal{O}(s-s_0)$ when $s\to s_0$ ($s$ denotes the curve parameter, and $s_0$ is an arbitrary value in the parameter range). Here $\lVert \mathrm{d}L(s)/\mathrm{d}s \rVert$ is required to be bounded and nonzero, thus reparameterizations of the curve parameter that would modify the leading-order scaling are excluded~\footnote{The scaling behavior may appear to change under a reparameterization of the curve $P(s)$, for example, $s^{\prime} = (s-s_{\text{EP}})^2$ or $s^{\prime} = (s-s_{\text{EP}})^{1/2}$ (where $P(s_{\text{EP}})$ is the EP). However, such reparameterizations would violate the requirement that $\lVert \mathrm{d}L(s)/\mathrm{d}s \rVert$ remain bounded and nonzero along the curve. This requirement arises from the validity of matrix perturbation theory; see Eq. (\ref{perturbation_of_A}), in which $A^{(1)}$ corresponds to $\mathrm{d}L(s)/\mathrm{d}s$ on the curve.}. In stark contrast, when the parameters migrate away from the EL, we observe $s^{1/2}$ scaling as $s\to 0$ in parameter curves (here $s_{\text{EP}}=0$ corresponds to the intersection point of the curve and the EL), indicating that the spectrum is highly sensitive to parameter perturbations that deviate from EL. Thus the spectral instability is anisotropic for EL.
%This contrasts sharply with the parameter migrating away from the EL, where at a fixed width we observe the $s^{1/2}$ scaling when $s\to 0$ in parameter curves (here $s_{\text{EP}}=0$ corresponds to the intersection point of the curve and the EL)

To investigate the topological structures of QNMs associated with this EL, we examine three closed curves $C_i(s),\,i=1,2,3$ in the $3$D parameter space:
\begin{eqnarray}
&&\Big(b_{1i}+a_{1i}\cos(4\pi s)\, ,b_{2i}+a_{2i}\sin(4\pi s)\, ,\nonumber\\
    &&b_{3i}+a_{3i}\sin(2\pi s+\theta_i)\Big)\, ,\quad s\in[0,1]\, ,
\end{eqnarray}
where $a_{ji},\,b_{ji}$ and $\theta_i$ are constant coefficients \footnote{These coefficients are explicitly given by: \\
$\begin{array}{llll}
    b_{11}=0.005\, ,&a_{11}=0.001\, ,&b_{21}=15.7\, ,&\\
    a_{21}=0.05\, ,&b_{31}=1\, ,&a_{31}=0.25\, ,&\theta_1=0\, ;\\
    b_{12}=0.005\, ,&a_{12}=0.001\, ,&b_{22}=15.7\, ,&\\
    a_{22}=0.05\, ,&b_{32}=1.125\, ,&a_{32}=0.125\, ,&\theta_2=-\pi/2\, ;\\
    b_{13}=0.005\, ,&a_{13}=-0.001\, ,&b_{23}=15.6\, ,&\\
    a_{23}=-0.04\, ,&b_{33}=1.125\, ,&a_{33}=0.125\, ,&\theta_3=0\, .
\end{array}$}. $C_1(s)$ and $C_2(s)$ encircle the EL twice, while $C_3(s)$ does not encircle it (see Fig. \ref{fig:paraloop}). All three curves exhibit counterclockwise rotation in the top view.

\begin{figure*}[htbp]
    \begin{minipage}[]{0.6\columnwidth}
        \includegraphics[width=\linewidth]{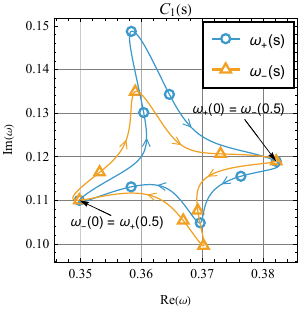}
    \end{minipage}\hspace{0.5cm}
    \begin{minipage}[]{0.6\columnwidth}
        {\includegraphics[width=\linewidth]{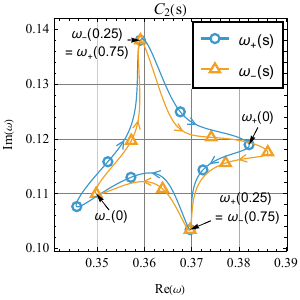}}
    \end{minipage}\hspace{0.5cm}
    \begin{minipage}[]{0.6\columnwidth}
        {\includegraphics[width=\linewidth]{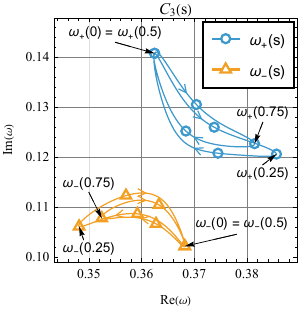}}
    \end{minipage}
    \caption{Trajectories of $\omega_{+}(s)$ and $\omega_{-}(s)$ along the three parameter-space curves $C_1(s)$, $C_2(s)$, and $C_3(s)$. The positions of $\omega_{\pm}$ are explicitly marked by $\circ$ and $\vartriangle$, respectively, at discrete parameter values $s = 0, 0.125, 0.25, \cdots, 1$.}
\label{fig:o1o2}
\end{figure*}

The two QNM spectra that coalesce at the EL is denoted as $\omega_+$ and $\omega_-$, and Fig. \ref{fig:o1o2} displays the trajectories of them for each curve. All trajectories evolve clockwise. For $C_1(s)$, the two QNM spectra exchange positions as $s$ increases from $0$ to $0.5$ and again from $0.5$ to $1$, returning to their initial positions at $s=1$. Similarly, for $C_2(s)$, exchange occurs between $s=0.25$ and $s=0.75$. In contrast, for $C_3(s)$, which does not encircle the EL, both $\omega_{+}(s)$ and $\omega_{-}(s)$ return to their initial positions without exchanging.

\begin{figure}[htbp]
    \centering
    \includegraphics[width=0.8\linewidth]{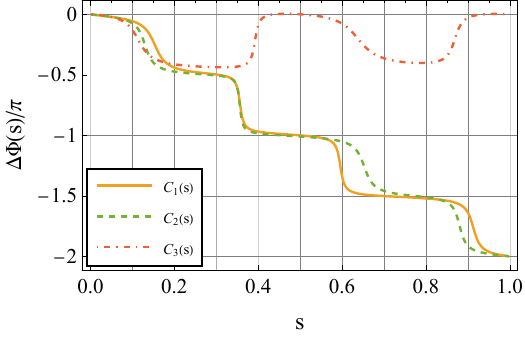}
    \caption{$\Delta\Phi(s)$ for $s$ varying from $0$ to $1$ for the three parameter-space curves $C_1(s), C_2(s)$, and $C_3(s)$.}
    \label{fig:argument}
\end{figure}
In non-Hermitian physics, the so-called vorticity can be introduced to describe the winding of each eigenvalues around a single EP~\cite{Leykam:2016qyj,Shen_2018,Ryu:2025gnn}. Its definition is
\begin{eqnarray}
    \nu(\Gamma)=-\frac{1}{2\pi}\oint_{\Gamma}\mathrm{d}\Phi\, ,
\end{eqnarray}
where $\Phi(s)=\arg\big(\omega_+(s)-\omega_-(s)\big)$, and $\Gamma$ is a closed loop that encircles a single EP once. For a second-order EP, the vorticity $\nu=\pm\frac{1}{2}$~\cite{Ryu:2025gnn}. To describe the winding of $\omega_+(s)$ and $\omega_-(s)$ in the complex plane, we plot the variation of $\Phi(s)$, i.e., $\Delta \Phi(s)=\Phi(s)-\Phi(0)$ for the three curves in Fig. \ref{fig:argument}. In our model, for a closed loop encircling the EL once, such as $s\in [0,0.5],\,s\in[0.5,1]$  for $C_1$ or $s\in[0.25,0.75]$ for $C_2$, each mode rotates $\pi$ relative to the other and their trajectories forms a closed loop, hence $\Delta\Phi(s)$ changes $-\pi$ and $\nu=1/2$. While, for a loop that do not encircle the EL like $s\in[0,0.5]\, \text{or}\, [0.5,1]$ for $C_3$, the trajectory of each mode forms a closed loop and there are no change of $\Delta\Phi(s)$ in one period, hence $\nu=0$.

Meanwhile, to study the phase evolution of these eigenvectors $v_{\pm}$ under adiabatic variation of the parameter along a closed loop, we investigate the Berry phase in non-Hermitian system defined in~\cite{Garrison1988,Dattoli_1990,Mostafazadeh:1999bjn,Liang:2015thd,Ryu:2025gnn}:
\begin{eqnarray}
    \gamma=\mathrm{i}\oint_{C(s)}\frac{\langle u_i(s),\partial_sv_i(s)\rangle}{\langle u_i(s),v_i(s)\rangle}\mathrm{d}s\, ,
\end{eqnarray}
where $i=\pm$ and $u_{\pm},v_{\pm}$ are left and right eigenvectors corresponding to $\omega_{\pm}$ respectively. The exact definitions of the left and right eigenvectors along with the inner product can be found in the Supplemental Material. Note that there is a gauge freedom: we can multiply $v_i(s)$ by an arbitrary smooth function $\alpha(s)$ with $\alpha(0)=\alpha(1)$, resulting an ambiguity in $\gamma$, namely
\begin{eqnarray}
    \gamma\rightarrow\gamma'=\gamma+    \mathrm{i}\oint\frac{\mathrm{d}\alpha}{\alpha}=\gamma+2k\pi\, ,\quad k\in\mathbb{Z}\, ,
\end{eqnarray}
where $k$ depends on the $\alpha(s)$ loop in the complex plane.

\begin{figure*}[]
    \begin{minipage}[]{0.65\columnwidth}
        \includegraphics[width=\linewidth]{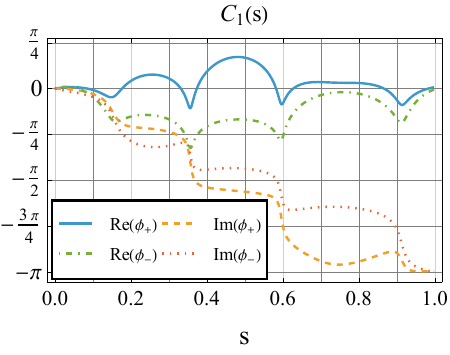}
    \end{minipage}\hspace{0.4cm}
    \begin{minipage}[]{0.65\columnwidth}
        {\includegraphics[width=\linewidth]{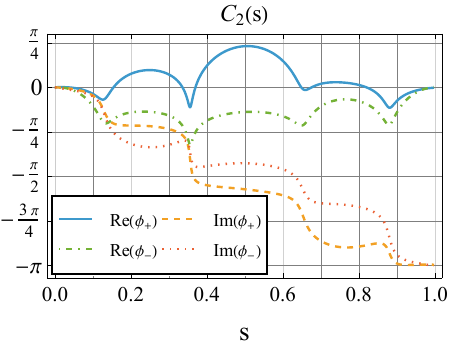}}
    \end{minipage}\hspace{0.4cm}
    \begin{minipage}[]{0.65\columnwidth}
        {\includegraphics[width=\linewidth]{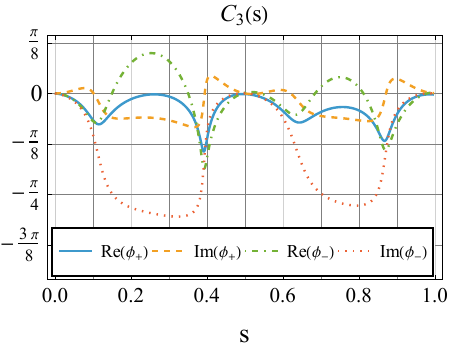}}
    \end{minipage}
    \caption{Real and imaginary parts of the integral \eqref{integral} evaluated for the two nearby modes on each of the three curves, where the blue, orange, green and red lines denote $\mathrm{Re}(\phi_+)\, ,\mathrm{Im}(\phi_+)\, ,\mathrm{Re}(\phi_-)$ and $\mathrm{Im}(\phi_-)$ respectively.}
\label{fig:gamma}
\end{figure*}

In practice, we use the spectral collocation method to obtain the finite-dimensional eigenvectors at each point of the parameter space. We fix the first element of the eigenvectors for gauge fixing by multiplying each eigenvector by a complex factor, which also ensures that the eigenvectors change continuously. The real and imaginary part of
\begin{eqnarray}\label{integral}
    \phi_i(s)\equiv\int_0^s\frac{\langle u_i(s^{\prime}),\partial_{s^{\prime}}v_i(s^{\prime})\rangle}{\langle u_i(s^{\prime}),v_i(s^{\prime})\rangle}\mathrm{d}s^{\prime}
\end{eqnarray}
for $s\in[0,1]$ are shown in Fig. \ref{fig:gamma}. Note that $\mathrm{Im}(\phi_i(1))=-\gamma$, while $\mathrm{Re}(\phi_i(1))$ vanishes theoretically. In Fig. \ref{fig:gamma}, at $s = 1$ the imaginary part is approximately $-\pi$ for $C_1$ and $C_2$, and approximately $0$ for $C_3$ at $s=0.5\, ,1$, while the real part vanishes in all cases. Thus, we confirm that the Berry phase is $\pi$ for a curve that encircles the EL twice and $0$ for a curve that do not encircle the EL, which is similar to the case of surrounding a single EP twice in a plane~\cite{Heiss:2012dx,Gao:2015cad,PhysRevE.69.056216,PhysRevLett.103.123003,Mailybaev:2005eet}.

\section{Pseudospectrum at exceptional points}\label{Pseudospectrum_at_EP}
The existence of EPs raises an important question about their spectral stabilities, and how their spectral stabilities differ from that of the non-EPs. To address this, we employ pseudospectrum analysis, which provides a powerful framework for investigating spectrum instability for the non-Hermitian systems~\cite{Destounis:2023ruj, Jaramillo:2020tuu,trefethen2020spectra}. For an operator $A$ and a given $\epsilon>0$, the $\epsilon$-pseudospectrum of the operator $A$ is defined as
\begin{eqnarray}\label{def of pseudo 1}
    \sigma_\epsilon(A)=\{z\in\mathbb{C}:\lVert(A-z I)^{-1}\rVert>\epsilon^{-1}\}\, ,
\end{eqnarray}
where $I$ denotes the identity operator. This definition proves most suitable for computational purposes. An equivalent yet more intuitive characterization is given by
\begin{eqnarray}\label{def of pseudo 2}
    \sigma_{\epsilon}(A) &=&\{z\in\mathbb{C}: \exists\delta A \, ,\text{with}\, \lVert\delta A\rVert<\epsilon\, ,\nonumber\\
    &&\text{such that}\, z\in\sigma(A+\delta A)\}\, ,
\end{eqnarray}
where $\sigma(A+\delta A)$ denotes the spectrum of $A+\delta A$. We also denote the boundary of $\sigma_\epsilon(A)$ as the $\epsilon$-contour.

To elucidate the pseudospectra at EPs, we examine the perturbation theory for linear operators. Given an operator $A^{(0)}$, we introduce a family of perturbed operators $A(\varkappa)$ on $A^{(0)}$,
\begin{equation}\label{perturbation_of_A}
    A(\varkappa) = A^{(0)} + \varkappa A^{(1)} \, , \qquad 0<\lVert A^{(1)}\rVert < \infty \, .
\end{equation}
We concentrate on finite-dimensional matrices in the follows. The corresponding eigenvalue problem becomes
\begin{eqnarray} 
    A(\varkappa)v_i(\varkappa)=\lambda_i(\varkappa)v_i(\varkappa)\, .
\end{eqnarray}
For systems with simple eigenvalues $\lambda_i$, standard perturbation theory yields the expansions
\begin{eqnarray}
    \lambda_i(\varkappa)&=&\lambda_i^{(0)}+\varkappa\lambda_i^{(1)}+\mathcal{O}(\varkappa^2)\, ,\label{eigenvalue_lambda}\\
    v_i(\varkappa)&=&v_i^{(0)}+\varkappa v_i^{(1)}+\mathcal{O}(\varkappa^2)\, .\label{eigenvector_v}
\end{eqnarray}

However, a parameter-dependent matrix is defective at its EP, and the conventional expansions \eqref{eigenvalue_lambda} and \eqref{eigenvector_v} of eigenvalues and eigenvectors are no longer valid. For a EP located at $\varkappa = 0$, consider a circular loop in the complex $\varkappa$-plane centered at the EP, starting from a point $\varkappa_0$. As $\varkappa$ traverses this loop once, a continuous tracing of the eigenvalues $\lambda_i(\varkappa)$ results in a permutation of the initial eigenvalues, $\lambda_i(\varkappa_0)$. This permutation decomposes into disjoint cycles~\cite{kato1976perturbation}:
\begin{align}
    \{\lambda_1, \lambda_2, \cdots, \lambda_{p_1}\},
    \{\lambda_{p_1+1}, \lambda_{p_1+2}, \cdots, \lambda_{p_1+p_2}\},\cdots ,
\end{align}
where each group undergoes a cyclic permutation after one encirclement in the $\varkappa$-plane. The number of elements $p_i$ in each group is defined as the period of the corresponding cycle. The permutation depends on the specific form of $A^{(1)}$, and we denote the period of the largest possible cycle as $q$.
%\footnote{A more rigorous description can be found in~\cite{1976Perturbation}.}

%For a $q$th-order exceptional point at $\lambda_i$ corresponding to $\varkappa=0$, if $\varkappa$ is restricted to a simply-connected subdomain $D$ excluding the exceptional point, the eigenvalues of $A(\varkappa)$ can be expressed as holomorphic functions~\cite{1976Perturbation} 
%\begin{eqnarray}
%    \lambda_h(\varkappa)\, ,\,h=1,\cdot\cdot\cdot,s\, .
%\end{eqnarray}
%All $s$ functions remain holomorphic on $D$ with $\lambda_h(\varkappa)\neq\lambda_k(\varkappa)$ for $h\neq k$. As $D$ undergoes continuous revolution around $\varkappa=0$, the $s$ eigenvalue functions $\lambda_h(\varkappa)$ experience permutations among themselves, grouping into cycles
%\begin{eqnarray}
%    \{\lambda_1,\lambda_2,\cdot\cdot\cdot,\lambda_{p_1}\},\{\lambda_{p_1+1},\lambda_{p_1+2},\cdot\cdot\cdot,\lambda_{p_1+p_2}\},\cdot\cdot\cdot\, ,
%\end{eqnarray}
%where each group undergoes a cyclic permutation by a revolution of $D$, and the number of elements in a cycle $p_i$ is called its period. 

The eigenvalues that form a cycle of period $p$ can be expanded as a Puiseux series  [c.f.~\cite{kato1976perturbation}]. For $\varkappa\neq0$, an eigenvalue $\lambda^{(0)}$ generally bifurcates into $p$ distinct branches, i.e.,
\begin{eqnarray}\label{lambda_of_kappa}
    \lambda_{h}(\varkappa)&=&\lambda^{(0)}+\lambda^{(1)}\Big(\mathrm{e}^{\frac{2\pi\mathrm{i}(h-1)}{p}}\varkappa^{1/p}\Big)^1\nonumber\\&&+\lambda^{(2)}\Big(\mathrm{e}^{\frac{2\pi\mathrm{i}(h-1)}{p}}\varkappa^{1/p}\Big)^2\nonumber+\mathcal{O}(\varkappa^{3/p})\, ,\nonumber\\ &&h=1,2,\cdots, p\, ,
\end{eqnarray}
which is also consistent with the cyclic permutation of period $p$. For small real $\varkappa$, the deviation $|\lambda_{h}(\varkappa)-\lambda^{(0)}|\propto\varkappa^{1/p}$ rather than $\propto\varkappa$, revealing the enhanced sensitivity of defective eigenvalues to perturbations compared to simple eigenvalues.

%For eigenvalue functions forming a cycle of period $p$, the eigenvalue functions can be expanded as Puiseux series. For small and real $\epsilon>0$ the eigenvalue generally bifurcates into $p$ distinct branches, i.e.,
%\begin{eqnarray}\label{lambda_of_kappa}
%    \lambda_{i}(\varkappa)&=&\lambda_i^{(0)}+\lambda_i^{(1)}(\omega^{(h-1)}\varkappa^{1/p})^1+\lambda_i^{(2)}(\omega^{(h-1)}\varkappa^{1/p})^2\nonumber\\
%    &&+\mathcal{O}(\varkappa^{3/p})\, ,
%\end{eqnarray}
%where $\omega=\exp(2\pi \mathrm{i}/p)$, $h=1,2,\cdot\cdot\cdot, p$. For small $\varkappa$, $|\lambda_{i}(\varkappa)-\lambda_i^{(0)}|\propto\varkappa^{1/p}$ rather than $\propto\varkappa$, revealing the enhanced sensitivity of defective eigenvalues to perturbations compared to simple eigenvalues.

Given a small $\epsilon>0$, consider the perturbation framework \eqref{perturbation_of_A} with real $\varkappa\leqslant\epsilon$ and arbitrary $A^{(1)}$ with $\lVert A^{(1)}\rVert=1$, the original $\lambda^{(0)}$ bifurcates into $\lambda_h$. As indicated by \eqref{lambda_of_kappa}, the deviation $\propto\varkappa^{1/p}$. Note that $\varkappa^{1/p}$ grows faster for larger $p$ when $\varkappa\to0$, thus the farthest possible deviation for all the $A^{(1)}$ is $\propto\epsilon^{1/q}$. On the other hand, in the context of pseudospectrum \eqref{def of pseudo 2}, an arbitrary perturbation $\delta A$ can be identified as $\delta A=\lVert\delta A\rVert\frac{\delta A}{\lVert\delta A\rVert}\equiv \lVert\delta A\rVert A^{(1)}=\varkappa A^{(1)}$. This leads to the conclusion that the $\epsilon$-contour, which marks the farthest deviation from the original eigenvalue, scales as $\propto \epsilon^{1/q}$. In case of a non-EP, where $q=1$, we recover the ordinary linear $\epsilon$ scaling behavior. We have the following theorem for the relationship between the largest period and the Jordan blocks of the matrix $A^{(0)}$, which is proved in the Supplemental Material.

\begin{theorem}\label{theorem1}
For a matrix perturbation problem $A(\varkappa)=A^{(0)}+\varkappa A^{(1)}$ with $\lVert A^{(1)}\rVert<\infty$, a degenerate eigenvalue $\lambda_i^{(0)}$ of $A^{(0)}$ bifurcates into several groups of eigenvalue $\{\lambda_1,\lambda_2,\cdots,\lambda_{p_1}\},\{\lambda_{p_1+1},\lambda_{p_1+2},\cdots,\lambda_{p_1+p_2}\},\cdots$, where each group undergoes a cyclic permutation when $\varkappa$ goes around $0$ in the complex plane. The largest possible period $q$ (or the largest possible number of the eigenvalues in a cycle) for all $A^{(1)}$ is just the index of the eigenvalue $\lambda_i^{(0)}$, i.e., the order of the largest Jordan block of the eigenvalue.
\end{theorem}
% This theorem becomes evident in special cases. Transforming $A$ into Jordan canonical form via similarity transformation
% \begin{eqnarray}
%     S^{-1}AS=J_A\, ,
% \end{eqnarray}
% then the perturbation problem for $A$ becomes equivalent to that of $J_A$ with modified perturbation matrix $\delta J_A=S^{-1}\delta AS$ and parameter $\epsilon_J$ ensuring $||\delta J_A||=1$. The $\epsilon_J$-pseudospectrum of $J_A$ then coincides with the $\epsilon$-pseudospectrum of $A$. If all the nonzero elements of $\delta J_A$ are in the same position of the Jordan blocks of $J_A$, the largest period $q$ is the index of the defective eigenvalue, i.e. the largest order of Jordan block of the eigenvalue. As for the general case where there may exist nonzero elements of $\delta J_A$ outside the position of the Jordan blocks, one can prove that the largest period is also the index of the defective eigenvalue [see Appendix\ref{proof}].

The following conclusions can be drawn from the above discussions: the $\epsilon$-contours of a matrix $A^{(0)}$ near a certain eigenvalue $\lambda^{(0)}$ scales as $\epsilon^{1/q}$ when $\epsilon\to0$, where $q$ is the order of the largest Jordan block of this eigenvalue. A toy model is provided in the Supplemental Material to illustrate this statement. 

\begin{figure*}[htbp]
    \begin{minipage}[]{\columnwidth}
        \includegraphics[width=\linewidth]{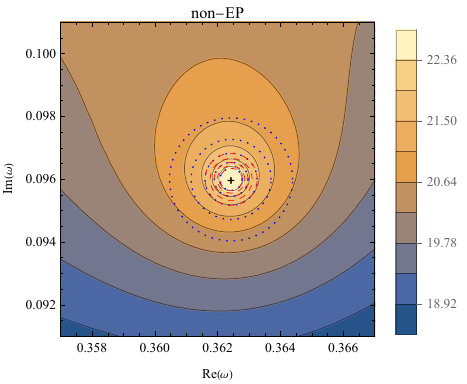}
        %\subcaption{\label{fig:Non-EP}Non-Exceptional point}
    \end{minipage}
    \hspace{0.4cm}
    \begin{minipage}[]{\columnwidth}
        {\includegraphics[width=\linewidth]{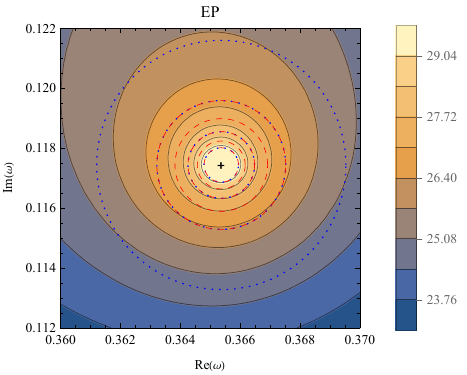}}
        %\subcaption{\label{fig:EP}Exceptional point}
    \end{minipage}
    \caption{Pseudospectrum of Regge-Wheeler potential with a bump modification near the fundamental mode at non-EP $\varepsilon=0.005, \,d=15, 2\sigma^2_0=1$  (left panel) and at the EP $\varepsilon=0.005083,\,d=15.6976, 2\sigma^2_0=1$ (right panel) by plotting $-\ln(\epsilon)$. The red dashed and blue dotted curves are circles centered at a QNM spectrum, with radii $B_1\epsilon^{1/2}$ and $B_2\epsilon$, respectively, where $\epsilon$ is chosen as the levels of the plotted contours. The constants $B_1$ and $B_2$ are determined by ensuring the smallest circle is inscribed in the innermost contour of the same $\epsilon$. This visually demonstrates the enhanced sensitivity at the EP, since in the EP case the red circles fit the outer contours much better than the blue ones, whereas in the non-EP case the blue ones fit better.}
\label{fig:pseudospectrum_EP_NP}
\end{figure*}

We now proceed to the computation of the black hole's pseudospectrum $\sigma_{\epsilon}(L/\mathrm{i})$ (see the Supplemental Material for the construction of $L$), noting that the choice of the norm significantly influences pseudospectrum's shape and structure. Fig. \ref{fig:pseudospectrum_EP_NP} displays pseudospectra near the fundamental mode at $\varepsilon=0.005$, $d=15$, $2\sigma_0^2=1$ (non-EP case) and at the exceptional point $\varepsilon=0.005083$, $d=15.6976$, $2\sigma_0^2=1$ (EP case). Our analysis reveals that inscribed circles with radii proportional to $\epsilon$ provide superior fitting for non-EP configurations, while radii scaling as $\epsilon^{1/2}$ better characterize EP cases, consistent with the above theoretical predictions.
%==================== Figure ====================%
%==================== Figure ====================

To further validate the scaling behavior of $\epsilon$-pseudospectrum contours, we analyze horizontal and vertical cross sections through the pseudospectrum structure (see the Supplemental Material). These cross-sectional analyses provide additional evidence for the characteristic $\epsilon^{1/2}$ scaling at EPs compared to the linear $\epsilon$ scaling away from EPs, reinforcing our theoretical predictions with comprehensive numerical verification.

%For a general perturbation with a particular norm, pseudospectrum analysis quantifies QNM spectrum \wlb{instability}, with our theoretical and numerical results \wlb{demonstrate} this enhanced \wlb{instability} at EPs.

The enhanced QNM spectra instability at EPs manifests clearly in the pseudospectrum structure. For sufficiently small $\epsilon$, the $\epsilon^{1/2}$ scaling demonstrates remarkably faster variation compared to linear $\epsilon$ scaling, indicating enhanced QNM spectrum instability when deviating from EPs in parameter space. The physical implications of these findings extend to gravitational wave physics and black hole spectroscopy. The $\epsilon^{1/2}$ scaling at second-order EPs implies that small perturbations could induce disproportionately large shifts in QNM spectra near EPs. The spectrum instability near EPs necessitates careful assessment of ringdown model reliability in gravitational wave analysis, as small systematic errors could be amplified in EP regions. Our results establish a framework for evaluating such spectrum stability across the black hole parameter space, with particular relevance for future high-precision gravitational wave observations.

%\z{A Gaussian bump with three variable parameters is added as a modification to the Regge-Wheeler potential, resulting a continuous line of EPs in the parameter space. The topological properties of this EL is investigated by calculating the vorticity and the Berry phase of loops encircling it. The spectral stability at EPs are studied through matrix perturbation theory and pesudospectrum.}

\section{Conclusions and discussion}\label{conclusions}
We find a continuous line of EPs (exceptional line, EL) in the $3$D parameter space $(\varepsilon,d,2\sigma_0^2)$ of the Gaussian bump-modified Regge-Wheeler potential. This EL exhibits rich topological characteristics: loops encircling the EL acquire the vorticity $\nu=\pm1/2$ and the Berry phase $\gamma=\pi$, while loops not encircling the EL yield $\nu=0$ and $\gamma=0$. Next, through matrix perturbation theory, we find that the $\epsilon$-pseudospectrum contour size scales as $\epsilon^{1/q}$ at an EP, where $q$ is the order of the largest Jordan block of the corresponding degenerate eigenvalue, in contrast to the linear $\epsilon$ scaling at non-EPs. Numerical computations at a second-order EP in the EL provide strong confirmation of this theoretical prediction, demonstrating $\epsilon^{1/2}$ scaling at second-order EPs compared to linear $\epsilon$ scaling away from EPs, which highlights the enhanced spectrum instability characteristic of non-Hermitian systems.

A crucial feature of the EL is its anisotropic spectral instability, which distinguishes it from isolated EPs. Moving along the EL direction in parameter space, the coalesced QNM spectra exhibit remarkable stability, whereas moving away from the EL reveals extreme spectral sensitivity. Such directional dependence has direct implications for black hole spectroscopy, implying that environmental perturbations directed parallel to the EL in parameter space induce milder shifts in the observed frequencies than those oriented differently. This anisotropic stability may prove valuable for identifying parameter regions in gravitational wave data analysis and for constructing more accurate waveform templates that account for the underlying non-Hermitian systems. Notably, in the parameterized analysis of QNMs,  when an EL is present, the fractional-power Puiseux expansion characteristic (similar to Eq. \eqref{lambda_of_kappa}) of an isolated EP is required only for parameter variations transverse to the EL; along or parallel to the EL direction, the expansion degenerates to an ordinary Taylor series with integer powers, reflecting the absence of the square-root branch point singularity in that direction.

The theoretical framework developed for matrix perturbation problems near EPs, particularly Theorem \ref{theorem1}, can be generalized to study eigenvalue sensitivity enhancement in non-Hermitian systems across various physical domains beyond black hole physics. Moreover, as the first application of pseudospectrum methods to investigate QNM spectrum stability near EPs in gravitational contexts, our approach provides a universal methodology that can be extended to study spectrum stability around EPs in other physical systems. Beyond the current context, similar EP phenomena may exist in perturbations of rotating black holes~\cite{Cavalcante:2024swt,Cavalcante:2024kmy,Cavalcante:2025abr}, wormholes, and exotic compact objects~\cite{Wu:2025wbp}. Although isolated EPs constitute a measure-zero set in parameter space, they can exert strong effects on parameter regions surrounding the EP through the change in topology they induce. The topological properties of these regions, including mode exchange when encircling EPs and the corresponding vorticity and Berry phase phenomena, warrant detailed investigation. Importantly, for gravitational wave observations, the heightened QNM spectral instability near EPs necessitates developing new waveform templates that account for this enhanced sensitivity, such as those proposed in~\cite{Yang:2025dbn,PanossoMacedo:2025xnf}, potentially improving the accuracy of black hole spectroscopy in future gravitational wave detectors. 
%======================================%
%<<<<<<<<<< Acknowledgement >>>>>>>>>>>%
%======================================%

\section*{Acknowledgement}
This work is supported in part by the National Key R\&D Program of China Grant No. 2022YFC2204603, by the National Natural Science Foundation of China with grants No.12475063, No. 12075232, No. 12247103. and No. 12505067.

\section*{Supplement Material}
\appendix
\section{The hyperboloidal framework}\label{hyperboloidal_framework}
The compact hyperboloidal coordinates $(\tau, \sigma, \theta, \varphi)$ related with original coordinate $(t, r, \theta, \varphi)$ are defined as follows
\begin{eqnarray}\label{compact_hyperboloidal_coordinates}
    t&=&\tau-h(\sigma)\, ,\nonumber\\
    r&=&\frac{r_{\text{h}}}{\sigma}\, ,
\end{eqnarray}
where $h(\sigma)$ is the height function. The coordinate $\sigma$ spans the interval $[0,1]$, with $\sigma=0$ corresponding to null infinity $\mathscr{I}^+$ and $\sigma=1$ to the event horizon $\mathscr{H}^+$. Following the approach in~\cite{PanossoMacedo:2023qzp}, we adopt the height function
\begin{eqnarray}
    h(\sigma)=2M\Big[-\frac{1}{\sigma}+\ln\sigma+\ln(1-\sigma)\Big]\, ,
\end{eqnarray}
and set $M=1$ for the unit. 

In these coordinates, the master perturbation equation \eqref{tdeq} transforms to
\begin{eqnarray}\label{maineq}
    \partial_\tau U=LU\, ,\quad
    L=
    \begin{pmatrix}
        0   & 1   \\
        L_1 & L_2
    \end{pmatrix}\, ,\quad
    U=\begin{pmatrix}
        \Psi \\
        \Pi
    \end{pmatrix}\, ,
\end{eqnarray}
with $\Pi\equiv\partial_{\tau}\Psi$. The differential operators $L_1$ and $L_2$ are given by
\begin{eqnarray}
    L_{1}&=&\frac{1}{w(\sigma)}\Big[\partial_{\sigma}(p(\sigma)\partial_{\sigma})-q(\sigma)\Big]\, ,\label{operator_L1}\\
    L_{2}&=&\frac{1}{w(\sigma)}\Big[2\gamma(\sigma)\partial_{\sigma}+\partial_{\sigma}\gamma(\sigma)\Big]\, ,\label{operator_L2}
\end{eqnarray}
where above functions read
\begin{eqnarray}\label{function_p_gamma_w_qx}
    p(\sigma)&=&\frac{\sigma^2f(r(\sigma))}{r_{\text{h}}}\, ,\quad \gamma(\sigma)=h^{\prime}(\sigma)p(\sigma)\, ,\nonumber\\
    w(\sigma)&=&\frac{1-\gamma^2(\sigma)}{p(\sigma)}\, ,\quad q(\sigma)=\frac{V(r(\sigma))}{p(\sigma)}\, ,
\end{eqnarray}
where $f(r)=1-2M/r$, $V(r)$ is the bump-modified potential in Eq. \eqref{tdeq} and the prime denotes differentiation with respect to $\sigma$. Consider the Fourier transform of $U(\tau,\sigma)$ with respect to time $\tau$, we have
\begin{eqnarray}
    U(\tau, \sigma)=\mathrm{e}^{\mathrm{i} \omega t}\tilde{U}(r_\star)=\mathrm{e}^{\mathrm{i}\omega\tau}U(\sigma)\, ,
\end{eqnarray}
where $U(\sigma)=\mathrm{e}^{-\mathrm{i}\omega h(\sigma)}\tilde{U}(r_\star)$ from Eqs. (\ref{compact_hyperboloidal_coordinates}). Then the problem of solving QNM spectra turns into an eigenvalue problem $LU=\mathrm{i}\omega U$. In practice, we apply the spectral collocation method and discretize $\sigma$ into Chebyshev-Lobatto grid with $N=300$. 

To support the concepts appeared in this work, including the vorticity, the Berry phase, the Petermann factors and the pseudospectrum of EPs, we give an introduction to the left and right eigenvectors along with the inner product and the norm. Consider a linear operator $A$ acting on a complex vector space equipped with inner product $\langle \cdot,\cdot\rangle$, and denote its adjoint as $A^{\dagger}$, defined through $\langle A^{\dagger}u,v\rangle=\langle u,Av\rangle$. The left $u_i$ and right $v_i$ eigenvectors satisfy respectively 
\begin{eqnarray}\label{lreigen}
    A^{\dagger}u_i=\Bar{\lambda}_iu_i\, ,\quad Av_i=\lambda_iv_i\, ,
\end{eqnarray}
where $\Bar{\lambda}_i$ is the complex conjugate of $\lambda_i$. We adopt the energy inner product 
\begin{eqnarray}\label{EnergyScalarProduct}
    &&\langle U_1, U_2\rangle_{\text{E}}=\left\langle\begin{pmatrix}
        \Psi_1 \\
        \Pi_1
    \end{pmatrix}, \begin{pmatrix}
        \Psi_2 \\
        \Pi_2
    \end{pmatrix}\right\rangle_{\text{E}}\nonumber\\
    &=&\frac{1}{2} \int_{0}^{1} \Big(w(\sigma)\bar{\Pi}_1 \Pi_2 + p(\sigma)  \partial_\sigma \bar{\Psi}_1\partial_\sigma\Psi_2 + q(\sigma)\bar{\Psi}_1 \Psi_2 \Big)\mathrm{d}\sigma\, ,\nonumber\\
\end{eqnarray}
where a bar represents the complex conjugate, and three functions $w(\sigma)$, $p(\sigma)$, $q(\sigma)$ are given by Eqs. (\ref{function_p_gamma_w_qx}). 
Therefore, a physically meaningful norm for the function $U(\sigma)$, namely the energy norm~\cite{Jaramillo:2020tuu, Gasperin:2021kfv, Besson:2024adi}, can be induced as
\begin{eqnarray}\label{energy_norm}
    \lVert U\rVert_{\text{E}}=\sqrt{\langle U, U\rangle_{\text{E}}}\, ,
\end{eqnarray}
which is chosen to implement the pseudospectrum.

Solving the eigenvalue problem, we identify EPs where the fundamental mode coincides with the first overtone and their eigenvectors coalesce. The model in~\cite{Yang:2025dbn} corresponds to the special case $2\sigma_0^2=1$, and for this case, we find an EP at $\varepsilon=0.005083$, $d=15.6976$ with nearly coincident spectra $\omega_{+}=0.36520+0.11743\mathrm{i}$ and $\omega_{-}=0.36554+0.11745\mathrm{i}$. Their eigenvectors also demonstrate near-perfect coalescence, confirming such parameter as a genuine EP. The phase rigidity $r_i$ and Petermann factor $K_i$ are defined as
\begin{eqnarray}\label{phase_rigidity_Petermann_factor}
    r_i=\frac{|\langle u_i,v_i\rangle|}{\sqrt{\langle u_i,u_i\rangle\langle v_i,v_i\rangle}}\equiv\frac{1}{\kappa_i}\, ,\quad K_i=\frac{1}{r_i^2}\equiv\kappa_i^2\, ,
\end{eqnarray}
where $u_i$ and $v_i$ are left and right eigenvectors defined in Eqs. \eqref{lreigen}. In Eqs. (\ref{phase_rigidity_Petermann_factor}),  $\kappa_i$ is the condition number, and the subscript $i$ refers to the index of eigenvalues. For normal operators, $\kappa_i=K_i=r_i=1$, whereas for non-normal operators $\kappa_i>1$, $K_i>1$, and $r_i<1$. Using the energy norm defined in Eq. \eqref{energy_norm}, it can be found that $r_+\simeq 2.03\times 10^{-10}$, $r_-\simeq 2.04\times 10^{-10}$ and $K_+\simeq 2.43\times 10^{19}$, $K_-\simeq 2.40\times 10^{19}$, indicating extreme non-normality around aforementioned EP.

\section{Proof of the Theorem \ref{theorem1}}\label{proof}
We first prove, for small real $\varkappa$, that the largest possible value of $p$ in the expansion given by Eq. \eqref{lambda_of_kappa} is equal to the index of the eigenvalue. Then, the equivalence between the period of the largest permutation cycle and the largest possible value of $p$ can be derived through complex analytical continuation from the expansion of real $\varkappa$ in Eq. \eqref{lambda_of_kappa}.

Consider a matrix $A(\varkappa)=A^{(0)}+\varkappa A^{(1)}$ and $A^{(1)}$ is bounded. we consider an eigenvalue $\lambda(\varkappa)$ and one of its eigenvectors $v(\varkappa)$ whose expansion are the same as \eqref{eigenvalue_lambda} and \eqref{eigenvector_v} respectively but the subscript $i$ is ignored. To begin with, the definition of root subspace or generalized eigenspace of a $n\times n$ matrix $A$ and its eigenvalue $\lambda$ is 
\begin{eqnarray}
    \mathcal{R}_{\lambda}(A):=\{ v\in \mathbb{C}^n| \,\exists k\in \mathbb{N},(A-\lambda I)^k v=0\}\, .
\end{eqnarray}

The smallest positive integer $\nu$ that satisfies
\begin{eqnarray}\label{def_index}
    (A^{(0)}-\lambda^{(0)} I)^{\nu} u=0\, ,\forall u\in\mathcal{R}_{\lambda^{(0)}}(A^{(0)})\, ,
\end{eqnarray}
from which $\nu$ is called the index of the eigenvalue $\lambda^{(0)}$, which is just the order of the largest Jordan block of it. We note that for every integer $k$ larger than the index, $(A^{(0)}-\lambda^{(0)} I)^k u=0\, ,\forall u\in\mathcal{R}_{\lambda^{(0)}}(A)$.  Noticing this identity for polynomials 
\begin{eqnarray}
    X^{\nu}-Y^{\nu}&=&(X-Y)(X^{\nu-1}+X^{\nu-2}Y\nonumber\\&&+\cdot\cdot\cdot+Y^{\nu-1})\, ,
\end{eqnarray}
for commutable matrices $A(\varkappa)-\lambda^{(0)} I$ and $(\lambda(\varkappa)-\lambda^{(0)})I$, we can obtain
\begin{eqnarray}\label{identity_of_A}
    &&(A(\varkappa)-\lambda^{(0)} I)^{\nu}-(\lambda(\varkappa)-\lambda^{(0)})^{\nu}I\nonumber\\&=&(A(\varkappa)-\lambda(\varkappa)I)\cdot Q(A(\varkappa),\lambda(\varkappa))\, ,
\end{eqnarray}
where $Q(A(\varkappa),\lambda(\varkappa))$ is a polynomial matrix of $A(\varkappa)$ and $\lambda(\varkappa)$ of order $\nu-1$.

The eigenvector $ v(\varkappa)$ corresponding to $\lambda(\varkappa)$ satisfies
\begin{eqnarray}\label{eigenvalue}
   ( A(\varkappa)-\lambda(\varkappa)I) v(\varkappa)=0\, .
\end{eqnarray}
Multiplying both side of the Eq. \eqref{identity_of_A} to $v(\varkappa)$, and we can find that the right hand side vanishes, thus
\begin{align}
    (A^{(0)}+\varkappa A^{(1)}-\lambda^{(0)} I)^{\nu} v(\varkappa)=(\lambda(\varkappa)-\lambda^{(0)})^{\nu} v(\varkappa)\, .
\end{align}
We expand the left hand side as Taylor series of $\varkappa$. As $v^{(0)}\in\mathcal{R}_{\lambda^{(0)}}(A^{(0)})$, the $\mathcal{O}(\varkappa^0)$ term $ (A^{(0)}-\lambda^{(0)} I)^{\nu}v^{(0)}=0$, therefore
\begin{align}
    C_1\varkappa v^{(0)}+C_2\varkappa v^{(1)}+\mathcal{O}(\varkappa^2)=(\lambda(\varkappa)-\lambda^{(0)})^{\nu} v(\varkappa)\, ,
\end{align}
in which $C_1$ and $C_2$ are bounded constant matrices. Taking the norm of the above equation and we can obtain
\begin{eqnarray}\label{neq1}
    &&|\lambda(\varkappa)-\lambda^{(0)}|^{\nu}\lVert v(\varkappa)\rVert\nonumber\\&=&\lVert C_1\varkappa v^{(0)}+C_2\varkappa v^{(1)}+\mathcal{O}(\varkappa^2)\rVert\nonumber\\
    &\leqslant&(\lVert C_1 v^{(0)}\rVert+\lVert C_2 v^{(1)}\rVert)\varkappa+\mathcal{O}(\varkappa^2)\, ,
\end{eqnarray}
where we set $\varkappa>0$ for simplicity. Divide both sides of \eqref{neq1} by $\lVert v(\varkappa)\rVert$, hence
\begin{align}
    |\lambda(\varkappa)-\lambda^{(0)}|^{\nu}&\leqslant \frac{\lVert C_1 v^{(0)}\rVert+\lVert C_2 v^{(1)}\rVert}{\lVert v(\varkappa) \rVert}\varkappa+\mathcal{O}(\varkappa^2)\nonumber\\
    &= \frac{\lVert C_1 v^{(0)}\rVert+\lVert C_2 v^{(1)}\rVert}{\lVert v^{(0)} \rVert}\varkappa+\mathcal{O}(\varkappa^2)\, .
\end{align}

%Therefore there exist a constant $C=(\lVert C_1 v^{(0)}\rVert+\lVert C_2 v^{(1)}\rVert)/\lVert v^{(0)}\rVert$ such that
%\begin{eqnarray}
   % |\lambda(\varkappa)-\lambda^{(0)}|^{\nu}\leqslant C\varkappa+\mathcal{O}(\varkappa^2)\, .
%\end{eqnarray}
Thus, for a small enough $\varkappa$, there exist a constant $K>0$ and
\begin{eqnarray}
     |\lambda(\varkappa)-\lambda^{(0)}|^{\nu}\leqslant K\varkappa \, ,
\end{eqnarray}
or equivalently
\begin{eqnarray}
    |\lambda(\varkappa)-\lambda^{(0)}|\leqslant K^{1/\nu}\varkappa^{1/\nu}\, .
\end{eqnarray}
This indicates that the largest possible $p$ in the expansion Eq. \eqref{lambda_of_kappa} is no greater than the index $\nu$. Moreover, there exists a kind of perturbation such that $p$ saturates the upper bound  $\nu$. Consider the Jordan decomposition of the matrix $A^{(0)}$, which contains a $\nu\times \nu$ Jordan block of eigenvalue $\lambda^{(0)}$
\begin{eqnarray}
   J_\nu = 
\begin{pmatrix}
\lambda^{(0)} & 1       &         &         & 0 \\
         &\, \lambda^{(0)} &\, \ddots &         &   \\
         &         &\, \ddots & \,\ddots      &   \\
         &         &         & \lambda^{(0)} & 1 \\
0        &         &         &         & \lambda^{(0)}
\end{pmatrix}\, ,
\end{eqnarray}
we can choose the perturbation matrix such that it perturbs $J_\nu$ as 
\begin{eqnarray}
   \varkappa \delta J_\nu = 
\begin{pmatrix}
    0&0&\cdots&0\\
    \vdots&\vdots&&\vdots\\
    0&0&\cdots&0\\
    \varkappa&0&\cdots&0 
\end{pmatrix}\, ,
\end{eqnarray}
then the eigenvalues of $J_\nu(\lambda)+\varkappa \delta J_\nu(\lambda)$ are $\lambda_h=\lambda^{(0)}+\varkappa^{1/\nu}\exp\left(\frac{2\pi\mathrm{i} (h-1)}{\nu}\right)$ with $h=1,2,\cdots,\nu$. After the complex analytical continuation on $\varkappa$ they are all in one cycle with period $\nu$.

Therefore, we have proved that for a fixed $A^{(0)}$ the largest period of eigenvalue $\lambda_i(\varkappa)$ for $A(\varkappa)=A^{(0)}+\varkappa A^{(1)}$ for all possible $A^{(1)}$ is the index of the eigenvalue $\lambda_i$ of $A^{(0)}$.

\section{toy models}\label{toy_model}
In this appendix, in order to show the essential differences of the pseudospectra between the non-EP and the EP (c.f. Theorem \ref{theorem1}). First, we focus on a toy model, namely a two-dimensional non-normal matrix (unless $x=1$) given by
\begin{eqnarray}\label{toymatrix}
    \textbf{A}(x)=\begin{pmatrix}
    x & 1\\
    x & -x
    \end{pmatrix}\, ,
\end{eqnarray}
where $x\in\mathbb{C}$ refers to the parameter of $\textbf{A}$.
%\z{\sout{ Note that $\textbf{A}(x)=\textbf{A}^{(0)}+x\textbf{A}^{(1)}$,with
% \begin{eqnarray}
%     \textbf{A}^{(0)}=\begin{pmatrix}
%     0 & 1\\
%     0 & 0
%     \end{pmatrix}\, ,
%     \textbf{A}^{(1)}=\begin{pmatrix}
%     1 & 0\\
%     1 & -1
%     \end{pmatrix}\, .
% \end{eqnarray}
% $\textbf{A}^{(0)}$ has only one second-order Jordan block corresponding to its eigenvalue $0$.}}
The eigenvalues of the matrix $\textbf{A}$ read $z_1=\sqrt{x+x^2}$ and $z_2=-\sqrt{x+x^2}$. It can be found that $x=0$ is a second-order exceptional point of $\textbf{A}$, and that $x=-1$ is another second-order exceptional point. The corresponding eigenvalue with the EPs is $z_0=0$ whatever $x=0$ or $x=-1$. At the EP $x=0$, the matrix $\textbf{A}(0)$ is given by
\begin{eqnarray}
    \textbf{A}(0)=\begin{pmatrix}
    0 & 1\\
    0 & 0
    \end{pmatrix}\, ,
 \end{eqnarray}
 which has only one second-order Jordan block corresponding to its eigenvalue $0$. In contrast, at non-EP the matrix $\textbf{A}(x)$ has two first-order Jordan block corresponding to $z_1$ and $z_2$ respectively.

Now, we will investigate the pseudospectrum of $\mathbf{A}(x)$, in which for simplicity, the $2$-norm is used in Eq. (\ref{def of pseudo 1}). Given $\epsilon>0$, one can analytically obtain the boundary of the $\epsilon$-pseudospectrum $\sigma_\epsilon(\textbf{A})$, which is given by the following equation associated with $z$:
\begin{eqnarray}\label{smallest_singularvalue}
    \epsilon=s_{\text{min}}(\textbf{A}-z\textbf{I})=\frac{\sqrt{3 x \bar{x}+2 z \bar{z}+1-\sqrt{\Delta}}}{\sqrt{2}}\, ,    
\end{eqnarray}
where the expression of $\Delta$ is
\begin{eqnarray}\label{Delta}
    \Delta&=&(\bar{x})^2 (5 x^2-4 x+4 z^2)\nonumber\\
    &&+2 \bar{x} (6 x z \bar{z}-2 x^2+x+2 z^2)\nonumber\\
    &&+4x(x+1)(\bar{z})^2+4z\bar{z}+1\, .
\end{eqnarray}

For the EP $x=0$, solving $z$ associated with $\epsilon$ from Eq. (\ref{smallest_singularvalue}) and Eq. (\ref{Delta}), one arrives at
\begin{eqnarray}
    |z-z_0|=\sqrt{\epsilon(1+\epsilon)}\, .
\end{eqnarray}
Therefore, the $\epsilon$-contours of $\sigma_\epsilon(\textbf{A})$ are a group of concentric circles centered at the origin whose radius are $\sqrt{\epsilon(1+\epsilon)}$. Similarly, for another EP $x=-1$, the $\epsilon$-contours are also a group of concentric circles centered at the origin with radius $\sqrt{\epsilon(2+\epsilon)}$. It is obvious that in both cases $|z-z_0|\propto \epsilon^{1/2}$ when $\epsilon\rightarrow 0$. Thus, $q=2$ and is just the order of the largest Jordan block of $\textbf{A}(0)$.

For the case that $x$ is real with $x\neq0$, there are two branches of equations of $\epsilon$-contour in polar coordinates, which are respectively
\begin{eqnarray}\label{rho_plus_rho_minus_Delta_1}
    \rho_{+}&=&\sqrt{x^2 \cos (2 \theta )+x \cos (2 \theta )+\epsilon ^2+ \sqrt{\Delta_1}}\, ,\label{rhop}\\
    \rho_{-}&=&\sqrt{x^2 \cos (2 \theta )+x \cos (2 \theta )+\epsilon ^2- \sqrt{\Delta_1}}\, ,\label{rhom}\\
    \Delta_1&=&\Big(x^2 \cos (2 \theta )+x \cos (2 \theta )+\epsilon ^2\Big)^2\nonumber\\
    &&-(x^4+2 x^3-3 x^2 \epsilon ^2+x^2+\epsilon ^4-\epsilon ^2)\, .
\end{eqnarray}

\begin{figure*}[htbp]
    \begin{minipage}[]{0.6\columnwidth}
        \includegraphics[width=\linewidth]{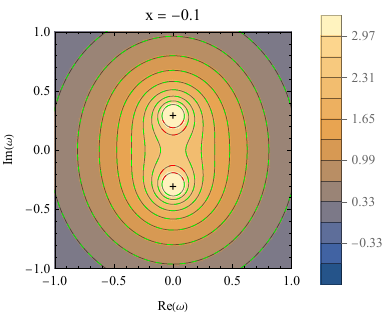}
    \end{minipage}\hspace{0.3cm}
    \begin{minipage}[]{0.6\columnwidth}
        {\includegraphics[width=\linewidth]{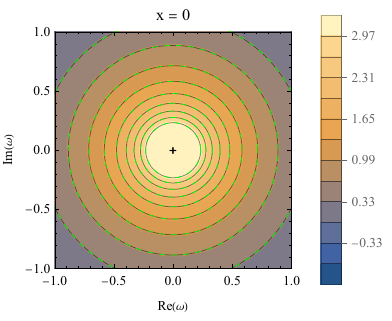}}
    \end{minipage}\hspace{0.3cm}
    \begin{minipage}[]{0.6\columnwidth}
        {\includegraphics[width=\linewidth]{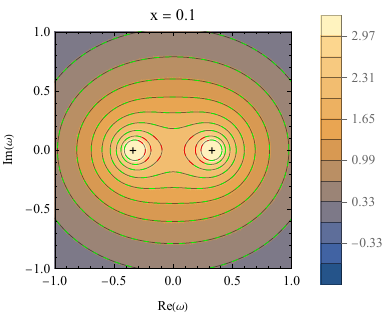}}
    \end{minipage}\hspace{0.3cm}
    \caption{Pseudospectra for the toy model [Eq. (\ref{toymatrix})] at $x=-0.1$, $0$, $1$. The dashed green lines are some of the $\rho_{+}$ branch curves and the dashed red lines are some of the $\rho_{-}$ branch curves. The symbols $+$ mark the eigenvalues of $\textbf{A}(x)$ in \eqref{toymatrix}.}
\label{fig:toymodel_pseudospectrum}
\end{figure*}

Then, we consider the process that $x$ varies within $(-1,1)$. When $-1<x<0$, $z_1$ and $z_2$ are pure imaginary. As $x$ increases to $0$, the distance between the two eigenvalues increases first and then decreases, and the two eigenvalues coincide at $x=0$. When $0<x<1$, as $x$ increases from $0$, the two eigenvalues bifurcates from $0$ to the real axis. In both cases $-1<x<0$ and $0<x<1$, there is a critical
\begin{eqnarray}
    \epsilon_c=\left|\frac{1}{2}(\sqrt{5 x^2+2 x+1}+x-1)\right|\, ,
\end{eqnarray}
such that for $\epsilon<\epsilon_c$ the $\epsilon$-contour is a set of closed curves around the two eigenvalues respectively, and for $\epsilon>\epsilon_c$ the contour is a curve that surrounds both two eigenvalues. This is caused by the fact that when $\epsilon<\epsilon_c$ the two branches $\rho_{+}$ and $\rho_{-}$ are all real and positive, and they together form a set of closed curves that surrounds the  two eigenvalue respectively, but when $\epsilon<\epsilon_c$ only $\rho_{+}$ is real and it forms a curve that surrounds the two eigenvalues. In order to confirm the above process, we plot $-\ln(\epsilon)$ together with some of the $\rho_{\pm}$ curves in Fig. \ref{fig:toymodel_pseudospectrum}. 

Now, we consider the scaling behavior of $\epsilon$-pseudospectrum contours for non-EP case. From Eq. \eqref{rhop} and Eq. \eqref{rhom}, we can find that for $x>0$ the $\epsilon$-contour surround $z_1$ satisfies
\begin{eqnarray}
    |z-z_1|&=&\sqrt{\rho^2+|z_1|^2-2\rho|z_1|\cos(\theta)}\nonumber\\
    &=& f(\theta,x)\epsilon+\mathcal{O}(\epsilon^2)\, ,
\end{eqnarray}
where $f(\theta, x)$ is a function of $\theta$ and $x$, and thus $|z-z_1|\propto \epsilon$ as $\epsilon\to0$. For the contour surround $z_2$ and the case $x<0$, or generally when the central eigenvalue is simple, we can similarly find that the radius of the $\epsilon$-contour proportional to $\epsilon$, which is consistent with the non-EP behavior mentioned in the main text. 

\begin{figure*}[htbp]
    \begin{minipage}[]{0.6\columnwidth}
        \includegraphics[width=\linewidth]{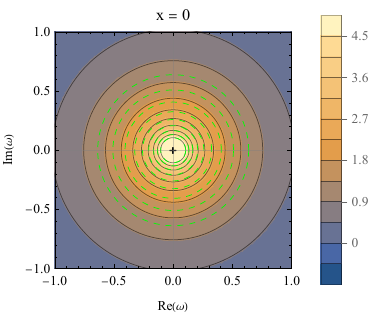}
    \end{minipage}\hspace{0.3cm}
    \begin{minipage}[]{0.6\columnwidth}
        {\includegraphics[width=\linewidth]{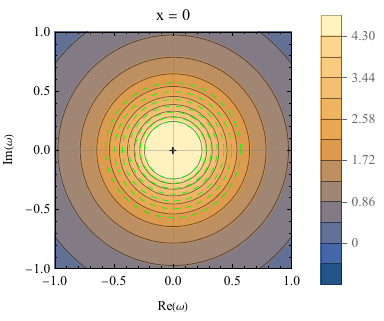}}
    \end{minipage}\hspace{0.3cm}
    \begin{minipage}[]{0.6\columnwidth}
        {\includegraphics[width=\linewidth]{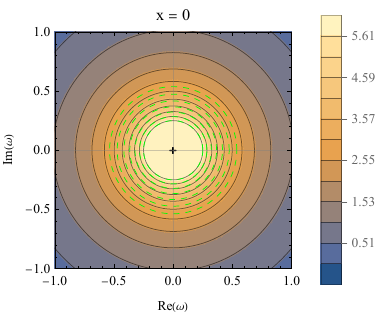}}
    \end{minipage}\hspace{0.3cm}
    \caption{Pseudospectra for $\textbf{A}_1(0)\, ,\textbf{A}_2(0)\, ,\text{and}\, \textbf{A}_3(0)$. We fit the smallest circle inscribed in the innermost contour, the dashed green lines are a group of circles centered in the eigenvalue $0$, and their radius are proportional to $\epsilon^{1/2},\,\epsilon^{1/3}\,\text{and}\,\epsilon^{1/4}$, respectively.}
\label{fig:toymodel_pseudospectrum2}
\end{figure*}

Finally, to test Theorem \ref{theorem1} for higher order EPs, we focus on three four-dimensional non-normal matrices $\textbf{A}_1(x),\,\textbf{A}_2(x),\,\textbf{A}_3(x)$  given by
\begin{align}\label{toymatrix2}
    \begin{pmatrix}
    x & 1 & 0 & 0\\
    0 & -x & 0 & 0\\
    0 &  0 & x & 0\\
    0 & 0 & 0 & 2
    \end{pmatrix},
    \begin{pmatrix}
    x & 1 & 0 & 0\\
    0 & -x & 1 & 0\\
    0 &  0 & x & 0\\
    0 & 0 & 0 & 2
    \end{pmatrix} ,
    \begin{pmatrix}
    x & 1 & 0 & 0\\
    0 & -x & 1 & 0\\
    0 &  0 & x & 1\\
    0 & 0 & 0 & x
    \end{pmatrix},
\end{align}
respectively. Within the same method of the two dimensional case in Eq. (\ref{toymatrix}), pseudospectra at an EP $x=0$ are depicted in Fig. \ref{fig:toymodel_pseudospectrum2}. As $\epsilon\to0$, the radius of the inscribed circles of the $\epsilon$-pseudospectrum boundaries proportional to $\epsilon^{1/q}$ with $q=2\, ,3 \, \text{and} \, 4$, which is consistent with the order of the largest Jordan block of $\textbf{A}_1(0)\, ,\textbf{A}_2(0)\, ,\textbf{A}_3(0)$ corresponding to the eigenvalue $0$.

\section{Horizontal and vertical sample points of pseudospectra for EP and non-EP configurations}\label{hvpseudo}
To quantitatively analyze the scaling behaviors of $\epsilon$-pseudospectrum contours, we examine horizontal and vertical sample points of the pseudospectra shown in Fig. \ref{fig:pseudospectrum_EP_NP}. Specifically, these sample points come from the left of and below the central spectrum (the fundamental mode for non-EP cases, and the coalesced mode for EP cases). 

\begin{figure*}[htbp]
    \begin{minipage}[]{\columnwidth}
        \includegraphics[width=\linewidth]{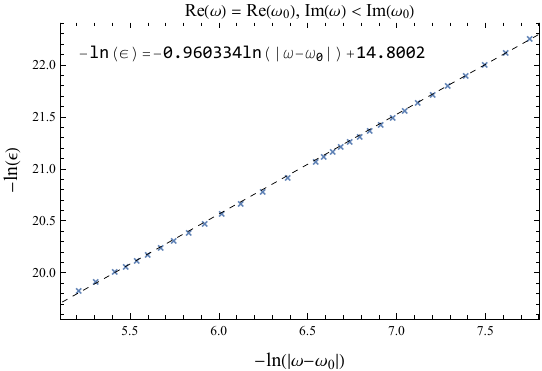}
    \end{minipage}
    \hspace{0.3cm}
    %\hfill
    \begin{minipage}[]{\columnwidth}
        {\includegraphics[width=\linewidth]{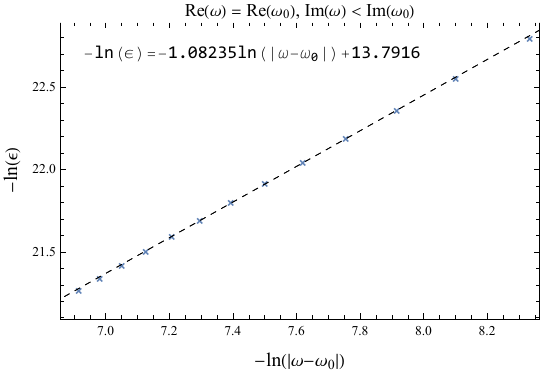}}
    \end{minipage}
    \caption{Horizontal (left) and vertical (right) sample points of pseudospectrum at $\varepsilon=0.005$, $d=15$, $2\sigma_0^2=1$ (non-EP case). The symbols $\times$ denote pseudospectrum data points, while the dashed lines show linear fits. The slope, approximately $1$, confirms the $\epsilon$ scaling for non-EP case.}
\label{fig:epomn}
\end{figure*}

\begin{figure*}[htbp]
    \begin{minipage}[]{\columnwidth}
        \includegraphics[width=\linewidth]{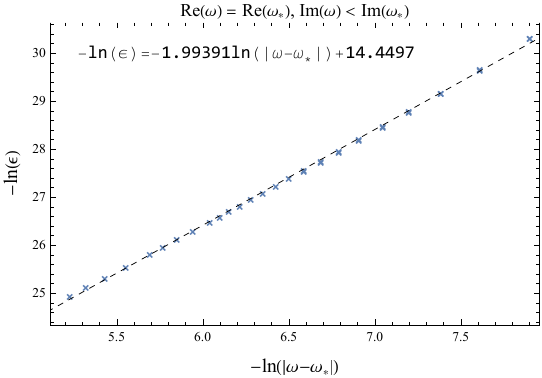}
    \end{minipage}
    \hspace{0.3cm}
    %\hfill
    \begin{minipage}[]{\columnwidth}
        {\includegraphics[width=\linewidth]{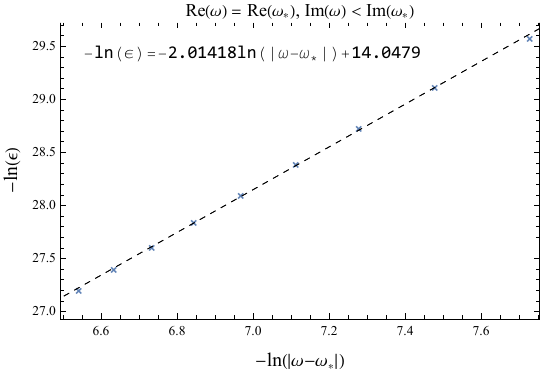}}
    \end{minipage}
    \caption{Horizontal (left) and vertical (right) sample points of pseudospectrum at the exceptional point $\varepsilon=0.005083$, $d=15.6976$, $2\sigma_0^2=1$ (EP case). The symbols $\times$ denote pseudospectrum data points, while the dashed lines show linear fits. The slope, approximately $2$, confirms the $\epsilon^{1/2}$ scaling predicted by our theoretical analysis for second-order EPs.}
\label{fig:epome}
\end{figure*}

We depict $-\ln(\epsilon)$ against $-\ln(|\omega-\omega_{0}|)$ for non-EP configuration in Fig. \ref{fig:epomn}, while $-\ln(|\omega-\omega_{\star}|)$ for EP configuration in Fig. \ref{fig:epome}, where $\omega_{\star}=(\omega_{+}+\omega_{-})/2$ provides an estimate of the coincident spectrum at the EP. The slope of these logarithmic plots directly reveals the scaling exponent: slopes approaching $1$ indicate linear $\epsilon$ scaling characteristic of non-EP cases, while slopes approaching $2$ confirm the $\epsilon^{1/2}$ scaling at EPs predicted by our theoretical analysis.

\section{The anisotropic spectral instability of EL}\label{anisotropic}
\begin{figure}[H]
    \centering
    \includegraphics[width=0.75\linewidth]{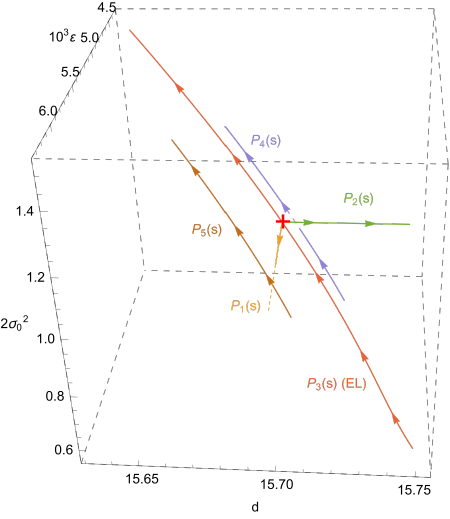}
    \caption{The curves $P_i(s)$ for $i=1,2,\cdots,5$. The ``+'' denotes the EP when $2\sigma_0^2=1$.}
    \label{fig:P1toP5}
\end{figure}

To verify the anisotropic spectral instability of the EL, we examine several parameter-space curves. The curves $P_1(s)$ and $P_2(s)$ move away from the EL, while $P_3(s)$ follows the EL itself, and $P_4(s)$ and $P_5(s)$ are curves parallel to the EL:
\begin{eqnarray}
    &&P_1(s): (\epsilon_{\star}+0.001s,d_{\star} ,1)  , s\in[0,0.5]\, ,\nonumber\\
    &&P_2(s): (\epsilon_{\star},d_{\star}+0.1s ,1)\, , s\in[0,0.5] \, ,\nonumber\\
    && P_3(s): \text{the EL}, s\equiv2\sigma_0^2\in[0.5,2]\, ,\nonumber\\
    && P_4(s): \text{the EL shifted by $-0.0002$ along $\epsilon$}\, ,\nonumber \\
    &&s\equiv2\sigma_0^2\in[0.8,1.2]\, ,\nonumber\\
   && P_5(s): \text{the EL shifted by $-0.02$ along $d$}\, , \nonumber\\
   &&s\equiv2\sigma_0^2\in[0.8,1.2]\, ,
\end{eqnarray}
where $\epsilon_{\star}=0.005083$ and $d_{\star}=15.6976$ is the parameter of EP at $2\sigma_0^2=1$, $s$ is the curve parameters. For $P_1$ and $P_2$, $s_{\text{EP}}=0$ corresponds to EP $(\varepsilon_{\star},d_{\star},1)$. We provide a diagram of these curves in Fig. \ref{fig:P1toP5}.

\begin{figure*}[htbp]
    \centering
    \subfigure[]{\includegraphics[width=0.3\linewidth]{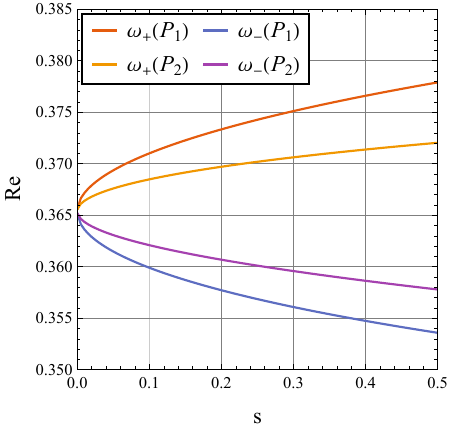}\label{fig:P12re}}  \hfill
    \subfigure[]{\includegraphics[width=0.3\linewidth]{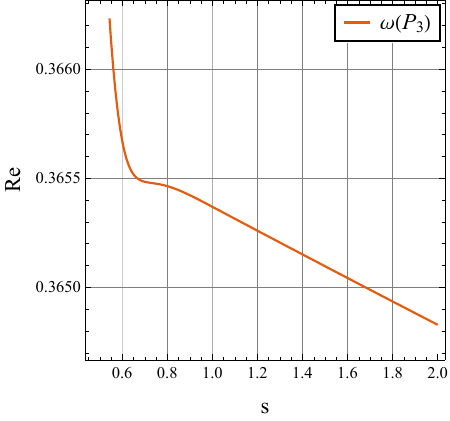}\label{fig:P3re}} \hfill
    \subfigure[]{\includegraphics[width=0.3\linewidth]{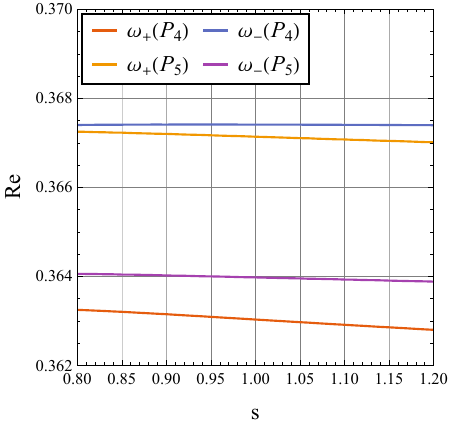}\label{fig:P45re}} \\
    \subfigure[]{\includegraphics[width=0.3\linewidth]{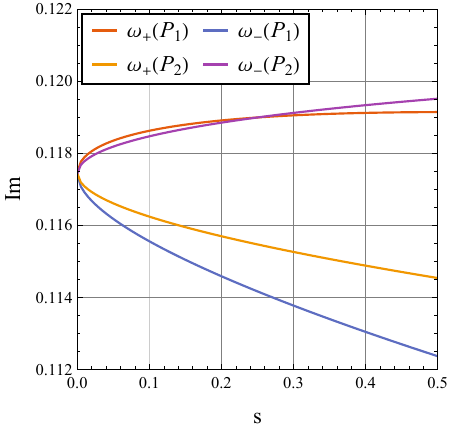}\label{fig:P12im}}  \hfill
    \subfigure[]{\includegraphics[width=0.3\linewidth]{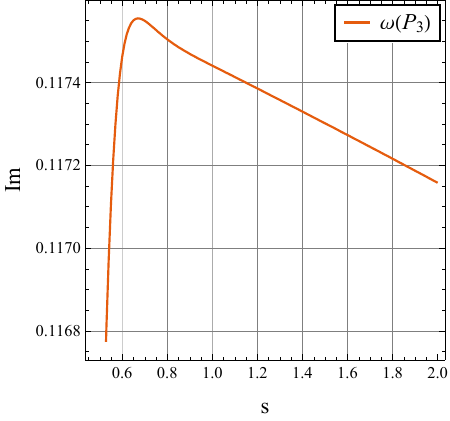}\label{fig:P3im}} \hfill
    \subfigure[]{\includegraphics[width=0.3\linewidth]{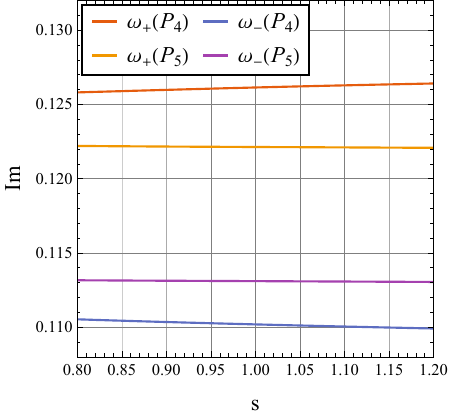}\label{fig:P45im}}
    \caption{Real and imaginary part of $\omega_+(s)$ and $\omega_-(s)$ along the parameter-space curves $P_i(s)$ for $i=1,2,\cdots,5$. For $P_3(s)$ (EL), the two modes coincide and are denoted by $\omega$.}
\label{fig:omegaovers}
\end{figure*}

Figs. \ref{fig:omegaovers} display the real and imaginary parts of $\omega_+(s)$ and $\omega_-(s)$ along the parameter-space curves $P_1(s)$ to $P_5(s)$. Due to numerical precision, the two modes do not perfectly coalesce at $s=0$ in Figs. \ref{fig:P12re} and \ref{fig:P12im}, resulting in a small but visible initial splitting. In Figs. \ref{fig:P3re} and \ref{fig:P3im}, where the curve follows the EL, the average $(\omega_{+}+\omega_{-})/2$ to represent the coalesced spectrum are shown, as the two modes are nearly degenerate along the exceptional line.

The key observation is that the drift of the QNM spectra scales differently between curves that deviate from the EL and those that remain on or parallel to it. For $P_1(s)$ and $P_2(s)$, which move away from the EL in the $\varepsilon$ and $d$ directions respectively, the spectra exhibit rapid variation with a leading-order scaling $\sim (s-s_{\text{EP}})^{1/2}$ when $s\to s_{\text{EP}}=0$, characteristic of the enhanced sensitivity at an EP. In contrast, along the EL itself ($P_3(s)$) and along curves shifted parallel to the EL ($P_4(s)$, $P_5(s)$), the spectra change slowly and the leading-order of $\omega(s)-\omega(s_0)$ is $\mathcal{O}(s-s_0)$ when $s\to s_0$ for an arbitrary $s_0$ in the parameter range. This stark contrast in scaling behavior directly demonstrates the anisotropic spectral instability of the EL: Moving away from the EL produce an $(s-s_{\text{EP}})^{1/2}$ spectral drift, thereby inducing the characteristic $\epsilon^{1/2}$ scaling of the pseudospectrum at parameter points on the EL and leading to enhanced spectral instability, while moving the parameters along or parallel to the EL leave the QNM spectra nearly unchanged to leading-order. This directional dependence, absent for isolated EPs, provides strong numerical evidence for the anisotropic nature of ELs.

\newpage
\bibliography{mainreference}
\bibliographystyle{apsrev4-1}

\end{document}